\documentstyle[12pt]{article}
\begin{document}
\begin{flushright}
Preprint IHEP 94-48 \\
31 March, 1994
\end{flushright}
\vspace*{1cm}
\begin{center}
{ \Large\bf Hadronic production of {\it\bf $B_c$}-mesons}
\end{center}

\vspace*{1cm}
\begin{center}
A.V.Berezhnoy\footnote{Moscow Institute of Physics and Technology},
 A.K.Likhoded\footnote{E-mail:LIKHODED@MX.IHEP.SU}, M.V.Shevlyagin
\end{center}
\begin{center}
Institute for High Energy Physics, Protvino, 142284, Russia
\end{center}
\vspace*{1cm}
{\bf Abstract}
On the basis of the exact formulas of QCD perturbation theory and parton model
 the hadronic production cross-sections for
$B_c({B_c}^*)$-mesons ($1^1S_0,1^3S_1,2^1S_0,2^3S_1$-states) are calculated.
The method used is the direct calculation of appropriate amplitudes
with the help of a FORTRAN program and subsequent Monte-Carlo integration
over the phase space and convolution with the structure functions.

\newpage

\section*{Introduction.}

Recently, there has been much interest in the study of $B$-mesons.
In the nearest future the $CP$-breaking parameters are planned
to be determined which suggests the production of $10^9-10^{10}$ $B$-mesons.
In these experiments the study of $B_c$-mesons, consisting of
$b$- and $\bar c$-quarks, will also be possible.
The $B_c$ production cross-section in comparison with that of $B$-mesons
 is suppressed approximately by three orders. Hence, having $10^6-10^7$
$B_c$-mesons one may hope to study in detail the properties of the $b\bar c$
system.

In a number of works [1-5] (see also more earlier works [6-10]),
devoted to the production of $B_c$-mesons and their exciting states
in $e^+e^-$ annihilation it is shown that the process of $B_c(B^*_c)$-meson
production at high energies can be represented as $b$-quark fragmentation
into $B_c$-meson. The analytical expression for the fragmentation
function $D_{b\rightarrow B_c}(z)$ and relative probability of
$\sim 10^{-3}$ is obtained, which allows one to relate the process of
$B_c$-meson production with the production of $b \bar b$ pair with
large invariant mass.

In $e^+e^-$ annihilation the production
cross-section of $B_c$-mesons is related with that of $b$-quarks in the
following way
\begin{equation}
\frac{d\sigma_{e^+e^-\to B_c(B^*_c){\bar b}c}}{dz}
= \sigma_{e^+e^- \to b \bar b}
\cdot D_{b\to B_c(B^*_c)}(z).
\end{equation}
where $z=|{\vec p}|/2\sqrt{s}$, $\vec p$ --- momentum of $B_c$-meson,
$\sqrt{s}$ --- total energy squared of $e^+e^-$ pair, $D_{b\to B_c}(z)$ and
$D_{b\to B^*_c}(z)$ --- fragmentation functions, while
$\sigma_{e^+e^- \to b{\bar b}}$ --- cross-section of the process
$e^+e^-\to b{\bar b}$, calculated in Born approximation.

One should note, however, that factorized form (1) in $e^+e^-$ annihilation
is realized when $M^2_{B_c}/s$ is small. Hence,  form (1) could be realized
in hadronic production only for large $p_{\bot}$ of $B_c$-meson. The final
answer in the case of
hadronic production of $B_c$-mesons turns out to be
more complex. First, in hadronic production the region
of small invariant masses dominates, where the asymptotical regime with
the cross-section factorization (1) is not realized yet. Second,
 for hadronic interactions the new type of diagram appears (called further
recombination diagrams) for which such factorization is absent.
The contribution of these diagrams, dominating at small invariant masses of the
$B_c \bar b c$ system, falls with the increase of this mass, but for all
studied
energies is substantial. As the result of exact numerical calculations of
matrix elements we will show that for reasonable values of the $B_c$-mesons
transverse momenta , produced at hadron colliders, the contributions of
both fragmentation and recombination types are significant.
For small transverse momenta the last one dominates.

The work is organized as follows. In the first Section the procedure
of matrix element calculation is presented. The behavior of
differential cross-sections of the subprocess $gg\rightarrow B_c\bar b c$
as a function of total gluon energy is discussed in the second Section.
 To obtain the cross-section of $B_c$-mesons
in hadronic collisions we convolute in Section 3
$\hat \sigma_{gg\rightarrow B_c\bar b c}$
with structure functions of gluons in initial hadrons.

\section{Calculation method.}

On the tree level the subprocess
$$g(p_1)+g(p_2) \rightarrow b(q_1) + \bar b(q_2)+ c(q_3) + \bar c(q_4)$$
is described by 38 Feynmann diagrams, that can be restored from
Fig. 1 and Table 1. As in [1,3,11] the amplitudes
can conveniently be calculated by  direct numerical multiplication
of $\gamma$-matrixes, spinors and polarization vectors.

Let us introduce the following notions:
\begin{displaymath}
\begin{array}{c}
q_1^2=q_2^2=m_1^2=m_2^2=m_b^2,\qquad q_3^2=q_4^2=m_3^2=m_4^2=m_c^2,  \\
q_{12}=-q_1-q_2,\qquad q_{34}=-q_3-q_4,\qquad p_{12}=p_1+p_2, \\
k_1=p_1+q_{12},\qquad k_2=p_1+q_{34}
\end{array}
\end{displaymath}

The three gluon vertexes are used in the form:
\begin{displaymath}
\begin{array}{c}
\Gamma ^\mu (p_a,p_b,\epsilon _a,\epsilon _b)
=-\Bigl( (2p_a+p_b) \cdot \epsilon _b \Bigr) \epsilon _a^{\mu}
+{(p_a-p_b)}^{\mu} \Bigl(\epsilon _a \cdot \epsilon _b \Bigr)
+\Bigl( (2p_b+p_a) \cdot \epsilon _a \Bigr) \epsilon _b^{\mu} , \\
\Gamma'^\mu (p_a,p_b,\epsilon _a,\epsilon _b)
=\Gamma ^\mu (p_a,p_b,\epsilon _a,\epsilon _b)/(p_a+p_b)^2
\end{array}
\end{displaymath}
Two quark currents are defined as follows:
 \begin{eqnarray}
J^{\mu}_{12} =
 \frac{\bar u (q_1) \gamma^{\mu} v(q_2)}{q^2_{12}} \qquad \qquad
J^{\mu}_{34} =
\frac{\bar u (q_3) \gamma^{\mu} v(q_4)}{q^2_{34}}. \nonumber
\end{eqnarray}
Besides, let us introduce subsidiary spinors:
\begin{eqnarray}
\bar u_{ij} = \bar u(q_i)\hat \epsilon _j\frac{ (\hat q_i -\hat p_j +m_i)}
{(q_i -p_j )^2 -m_i^2} \qquad \qquad \quad
v_{ij} = \frac{(\hat p_j -\hat q_i +m_i)}
{(p_j -q_i )^2 -m_i^2)}\hat \epsilon _j v(q_i ) \nonumber \\
\nonumber \\
\nonumber \\
\bar u_{1,34} = \frac{ \bar u(q_1){\hat J}_{34}(\hat q_1 -\hat q_{34} +m_1)}
{(q_1 -q_{34} )^2 -m_1^2 } \qquad \qquad
\bar u_{3,12} = \frac{ \bar u(q_3){\hat J}_{12}(\hat q_3 -\hat q_{12} +m_3)}
{(q_3 -q_{12} )^2 -m_3^2 }  \nonumber \\
\nonumber \\
v_{2,34} = \frac{(\hat q_{34} -\hat q_2 +m_2){\hat J}_{34} v(q_2 )}
{(q_{34} -q_2 )^2 -m_2^2}  \qquad \qquad
v_{4,12}  =  \frac{(\hat q_{12} -\hat q_4 +m_4){\hat J}_{12} v(q_4 )}
{(q_{12} -q_4 )^2 -m_4^2}. \nonumber
\end{eqnarray}
Gluon polarization vectors are used in the following way:
 $$\epsilon_1=\epsilon_1(p_1) \qquad \epsilon_2=\epsilon_2(p_2).$$
The  matrix element squared are summed over  orthogonal gluon states:
$$\epsilon '=(0,1,0,0) \  \ \epsilon ''=(0,0,1,0).$$
Note, that ${\epsilon '}^2 ={\epsilon ''}^2 = -1$, and $p \cdot \epsilon=0$,
where $p$ - gluon momentum. For colour indexes we use notions:

upper index $g_1,g_2$--gluons;

lower index $q_1$--$b$-quark;

lower index $q_2$--$\bar b$-quark;

lower index $q_3$--$c$-quark;

lower index $q_4$--$\bar c$-quark.

The contributions from appropriate diagrams ( see Fig. 1 and Tab. 1 )
 will have the following forms
(colour coefficients are surrounded by braces):

\newcommand{\F}[6]{f^{{#1_#2}{#3_#4}{#5_#6}}}
\newcommand{\T}[6]{t^{#1_#2}_{{#3_#4}{#5_#6}}}

\begin{displaymath}
\begin{array}{l}
M_1=\Bigl( \Gamma(p_2,q_{34},\epsilon_2,J_{34})\cdot
\Gamma'(p_1,q_{12},\epsilon_1,J_{12}) \Bigr)
\cdot\{\F {k}{1} {g}{2} {k}{2} \F {k}{3} {k}{2} {g}{1}
        \T {k}{1} {q}{3} {q}{4} \T {k}{3} {q}{1} {q}{2}\}
\\
M_2=\Bigl( \Gamma'(p_1,q_{34},\epsilon_1,J_{34})\cdot
\Gamma(p_2,q_{12},\epsilon_2,J_{12}) \Bigr)
\cdot\{\F {k}{1} {g}{2} {k}{2} \F {k}{3} {k}{2} {g}{1}
       \T {k}{3} {q}{3} {q}{4} \T {k}{1} {q}{1} {q}{2}\}
\\
M_3=\Bigl( \Gamma'(p_1,p_2,\epsilon_1,\epsilon_2)\cdot
\Gamma(q_{12},q_{34},J_{12},J_{34}) \Bigr)
\cdot\{ -\F {k}{1} {g}{2} {g}{1} \F {k}{2} {k}{1} {k}{3}
         \T {k}{2} {q}{1} {q}{2} \T {k}{3} {q}{3} {q}{4}\}
\\
M_4=J_{12}^\alpha J_{34}^\beta \Bigl( \epsilon_{1\beta}\epsilon_{2\alpha}-
(\epsilon_1\cdot\epsilon_2)g_{\alpha\beta}\Bigr)
\cdot\{ \F {k}{1} {g}{2} {k}{2} \F {k}{2} {k}{3} {g}{1}
         \T {k}{3} {q}{1} {q}{2} \T {k}{1} {q}{3} {q}{4}\}
\\
M_5=J_{12}^\alpha J_{34}^\beta\Bigl( (\epsilon_1\cdot\epsilon_2)g_{\alpha\beta}
-\epsilon_{1\alpha}\epsilon_{2\beta}\Bigr)
\cdot\{ \F {g}{2} {k}{1} {k}{2} \F {k}{2} {k}{3} {g}{1}
         \T {k}{1} {q}{1} {q}{2} \T {k}{3} {q}{3} {q}{4}\}
\\
M_6=J_{12}^\alpha J_{34}^\beta \Bigl(\epsilon_{1\alpha}\epsilon_{2\beta}
-\epsilon_{1\beta}\epsilon_{2\alpha}\Bigr)
\cdot\{ \F {k}{1} {k}{2} {k}{3} \F {k}{3} {g}{2} {g}{1}
         \T {k}{1} {q}{1} {q}{2} \T {k}{2} {q}{3} {q}{4}\}
\\
M_7=\bar u_{11}\hat\Gamma'(p_2,q_{34},\epsilon_2,J_{34})v(q_2)
\cdot\{ i\F {k}{1} {k}{2} {g}{2} \T {k}{1} {q}{3} {q}{4}
         \T {g}{1} {q}{1} {l}{1} \T {k}{2} {l}{1} {q}{2}\}
\\
M_8=\bar u(q_1)\hat\Gamma'(p_2,q_{34},\epsilon_2,J_{34})v_{21}
\cdot\{ i\F {k}{1} {k}{2} {g}{2} \T {k}{1} {q}{3} {q}{4}
         \T {k}{2} {q}{1} {l}{1} \T {g}{1} {l}{1} {q}{2}\}
\\
M_9=\bar u_{31}\hat\Gamma'(p_2,q_{12},\epsilon_2,J_{12})v(q_4)
\cdot\{ i\F {k}{1} {k}{2} {g}{2} \T {k}{1} {q}{1} {q}{2}
         \T {g}{1} {q}{3} {l}{1} \T {k}{2} {l}{1} {q}{4}\}
\\
M_{10}=\bar u(q_3)\hat\Gamma'(p_2,q_{12},\epsilon_2,J_{12})v_{41}
\cdot\{ i\F {k}{1} {k}{2} {g}{2} \T {k}{1} {q}{1} {q}{2}
         \T {k}{2} {q}{3} {l}{1} \T {g}{1} {l}{1} {q}{4}\}
\\
M_{11}=\bar u(q_3)\hat\Gamma'(p_1,q_{12},\epsilon_1,J_{12})v_{42}
\cdot\{ i\F {k}{1} {k}{2} {g}{1} \T {k}{1} {q}{1} {q}{2}
         \T {k}{2} {q}{3} {l}{1} \T {g}{2} {l}{1} {q}{4}\}
\\
M_{12}=\bar u_{32}\hat\Gamma'(p_1,q_{12},\epsilon_1,J_{12})v(q_4)
\cdot\{ i\F {k}{1} {k}{2} {g}{1} \T {k}{1} {q}{1} {q}{2}
         \T {g}{2} {q}{3} {l}{1} \T {k}{2} {l}{1} {q}{4}\}
\\
\end{array}
\end{displaymath}
\begin{displaymath}
\begin{array}{l}
M_{13}=\bar u(q_1)\hat\Gamma'(p_1,q_{34},\epsilon_1,J_{34})v_{22}
\cdot\{ i\F {k}{1} {k}{2} {g}{1} \T {k}{1} {q}{3} {q}{4}
         \T {k}{2} {q}{1} {l}{1} \T {g}{2} {l}{1} {q}{2}\}
\\
M_{14}=\bar u_{12}\hat\Gamma'(p_1,q_{34},\epsilon_1,J_{34})v(q_2)
\cdot\{ i\F {k}{1} {k}{2} {g}{1} \T {k}{1} {q}{3} {q}{4}
         \T {g}{2} {q}{1} {l}{1} \T {k}{2} {l}{1} {q}{2}\}
\\
M_{15}=\bar u_{11}\gamma^\alpha v(q_2)\bar u(q_3)\gamma_\alpha v_{42}/k_1^2
\cdot\{ \T {g}{1} {q}{1} {l}{1} \T {k}{1} {l}{1} {q}{2}
        \T {k}{1} {q}{3} {l}{2} \T {g}{2} {l}{2} {q}{4}\}
\\
M_{16}=\bar u(q_1)\gamma^\alpha v_{21}\bar u(q_3)\gamma_\alpha v_{42}/k_1^2
\cdot\{ \T {k}{1} {q}{1} {l}{1} \T {g}{1} {l}{1} {q}{2}
        \T {k}{1} {q}{3} {l}{2} \T {g}{2} {l}{2} {q}{4}\}
\\
M_{17}=\bar u_{11}\gamma^\alpha v(q_2)\bar u_{32}\gamma_\alpha v(q_4)/k_1^2
\cdot\{ \T {g}{1} {q}{1} {l}{1} \T {k}{1} {l}{1} {q}{2}
        \T {g}{2} {q}{3} {l}{2} \T {k}{1} {l}{2} {q}{4}\}
\\
M_{18}=\bar u(q_1)\gamma^\alpha v_{21}\bar u_{32}\gamma_\alpha v(q_4)/k_1^2
\cdot\{ \T {k}{1} {q}{1} {l}{1} \T {g}{1} {l}{1} {q}{2}
        \T {g}{2} {q}{3} {l}{2} \T {k}{1} {l}{2} {q}{4}\}
\\
M_{19}=\bar u_{31}\gamma^\alpha v(q_4)\bar u(q_1)\gamma_\alpha v_{22}/k_2^2
\cdot\{ \T {k}{1} {q}{1} {l}{1} \T {g}{2} {l}{1} {q}{2}
        \T {g}{1} {q}{3} {l}{2} \T {k}{1} {l}{2} {q}{4}\}
\\
M_{20}=\bar u(q_3)\gamma^\alpha v_{41}\bar u(q_1)\gamma_\alpha v_{22}/k_2^2
\cdot\{ \T {k}{1} {q}{1} {l}{1} \T {g}{2} {l}{1} {q}{2}
        \T {k}{1} {q}{3} {l}{2} \T {g}{1} {l}{2} {q}{4}\}
\\
M_{21}=\bar u_{31}\gamma^\alpha v(q_4)\bar u_{12}\gamma_\alpha v(q_2)/k_2^2
\cdot\{ \T {g}{2} {q}{1} {l}{1} \T {k}{1} {l}{1} {q}{2}
        \T {g}{1} {q}{3} {l}{2} \T {k}{1} {l}{2} {q}{4}\}
\\
M_{22}=\bar u(q_3)\gamma^\alpha v_{41}\bar u_{12}\gamma_\alpha v(q_2)/k_2^2
\cdot\{ \T {g}{2} {q}{1} {l}{1} \T {k}{1} {l}{1} {q}{2}
        \T {k}{1} {q}{3} {l}{2} \T {g}{1} {l}{2} {q}{4}\}
\\
M_{23}=\bar u_{1,34}\hat\epsilon_1 v_{22}
\cdot\{ \T {k}{1} {q}{1} {l}{1} \T {g}{1} {l}{1} {l}{2}
        \T {g}{2} {l}{2} {q}{2} \T {k}{1} {q}{3} {q}{4}\}
\\
M_{24}=\bar u_{11}\hat J_{34} v_{22}
\cdot\{ \T {g}{1} {q}{1} {l}{1} \T {k}{1} {l}{1} {l}{2}
        \T {g}{2} {l}{2} {q}{2} \T {k}{1} {q}{3} {q}{4}\}
\\
M_{25}=\bar u_{11}\hat\epsilon_2 v_{2,34}
\cdot\{ \T {g}{1} {q}{1} {l}{1} \T {g}{2} {l}{1} {l}{2}
        \T {k}{1} {l}{2} {q}{2} \T {k}{1} {q}{3} {q}{4}\}
\\
M_{26}=\bar u_{1,34}\hat\epsilon_2 v_{21}
\cdot\{ \T {k}{1} {q}{1} {l}{1} \T {g}{2} {l}{1} {l}{2}
        \T {g}{1} {l}{2} {q}{2} \T {k}{1} {q}{3} {q}{4}\}
\\
M_{27}=\bar u_{12}\hat J_{34} v_{21}
\cdot\{ \T {g}{2} {q}{1} {l}{1} \T {k}{1} {l}{1} {l}{2}
        \T {g}{1} {l}{2} {q}{2} \T {k}{1} {q}{3} {q}{4}\}
\\
M_{28}=\bar u_{12}\hat\epsilon_1 v_{2,34}
\cdot\{ \T {g}{2} {q}{1} {l}{1} \T {g}{1} {l}{1} {l}{2}
        \T {k}{1} {l}{2} {q}{2} \T {k}{1} {q}{3} {q}{4}\}
\\
M_{29}=\bar u_{3,12}\hat\epsilon_1 v_{42}
\cdot\{ \T {k}{1} {q}{3} {l}{1} \T {g}{1} {l}{1} {l}{2}
        \T {g}{2} {l}{2} {q}{4} \T {k}{1} {q}{1} {q}{2}\}
\\
M_{30}=\bar u_{31}\hat J_{12} v_{42}
\cdot\{ \T {g}{1} {q}{3} {l}{1} \T {k}{1} {l}{1} {l}{2}
        \T {g}{2} {l}{2} {q}{4} \T {k}{1} {q}{1} {q}{2}\}
\\
M_{31}=\bar u_{31}\hat\epsilon_2 v_{4,12}
\cdot\{ \T {g}{1} {q}{3} {l}{1} \T {g}{2} {l}{1} {l}{2}
        \T {k}{1} {l}{2} {q}{4} \T {k}{1} {q}{1} {q}{2}\}
\\
M_{32}=\bar u_{3,12}\hat\epsilon_2 v_{41}
\cdot\{ \T {k}{1} {q}{3} {l}{1} \T {g}{2} {l}{1} {l}{2}
        \T {g}{1} {l}{2} {q}{4} \T {k}{1} {q}{1} {q}{2}\}
\\
M_{33}=\bar u_{32}\hat J_{12} v_{41}
\cdot\{ \T {g}{2} {q}{3} {l}{1} \T {k}{1} {l}{1} {l}{2}
        \T {g}{1} {l}{2} {q}{4} \T {k}{1} {q}{1} {q}{2}\}
\\
M_{34}=\bar u_{32}\hat\epsilon_1 v_{4,12}
\cdot\{ \T {g}{2} {q}{3} {l}{1} \T {g}{1} {l}{1} {l}{2}
         \T {k}{1} {l}{2} {q}{4} \T {k}{1} {q}{1} {q}{2}\}
\\
M_{35}=\bar u_{1,34}\hat\Gamma' (p_1,p_2,\epsilon_1,\epsilon_2)v(q_2)
\cdot\{ -i\F {k}{1} {g}{2} {g}{1} \T {k}{2} {q}{1} {l}{1}
          \T {k}{1} {l}{1} {q}{2} \T {k}{2} {q}{3} {q}{4}\}
\\
M_{36}=\bar u(q_1)\hat\Gamma' (p_1,p_2,\epsilon_1,\epsilon_2)v_{2,34}
\cdot\{ -i\F {k}{1} {g}{2} {g}{1} \T {k}{1} {q}{1} {l}{1}
          \T {k}{2} {l}{1} {q}{2} \T {k}{2} {q}{3} {q}{4}\}
\\
M_{37}=\bar u_{3,12}\hat\Gamma' (p_1,p_2,\epsilon_1,\epsilon_2)v(q_4)
\cdot\{ -i\F {k}{1} {g}{2} {g}{1} \T {k}{2} {q}{3} {l}{1}
          \T {k}{1} {l}{1} {q}{4} \T {k}{2} {q}{1} {q}{2}\}
\\
M_{38}=\bar u(q_3)\hat\Gamma' (p_1,p_2,\epsilon_1,\epsilon_2)v_{4,12}
\cdot\{ -i\F {k}{1} {g}{2} {g}{1} \T {k}{1} {q}{3} {l}{1}
          \T {k}{2} {l}{1} {q}{4} \T {k}{2} {q}{1} {q}{2}\}
\\
\end{array}
\end{displaymath}

One can represent the matrix element of the given  subprocess in the following
form:
$$ M_{gg}=\sum_{\alpha=1}^{38}M_{\alpha}
=\sum_{\alpha =1}^{38}C_{\alpha}\cdot \tilde M_{\alpha}, $$
where $\tilde M_{\alpha}$ is a spinor part of the contribution of the
diagram with
number $\alpha$, $C_{\alpha}$--colour part and
$M_{\alpha}=C_{\alpha}\cdot \tilde M_{\alpha}$ is the total contribution
of the appropriate diagram.
\begin{table}[t]
\caption{The correspondence between the
diagrams of Fig. 1 and the
contributions $M_{\alpha}$ in the total four free quarks gluonic production
amplitudes.}
\begin{center}
\small
\begin{tabular}{|c|c|c|c|c|c|} \hline
 $ \alpha  $   & the diagram  & a & b &     1      &     2      \\ \hline
       1       &    1         & 1 & 2 & $b \bar b$ & $c \bar c$ \\ \hline
       2       &    1         & 2 & 1 & $b \bar b$ & $c \bar c$ \\ \hline
       3       &    2         & 1 & 2 & $b \bar b$ & $c \bar c$ \\ \hline
       4       &    11        & 1 & 2 & $b \bar b$ & $c \bar c$ \\ \hline
       5       &    11        & 1 & 2 & $b \bar b$ & $c \bar c$ \\ \hline
       6       &    11        & 1 & 2 & $b \bar b$ & $c \bar c$ \\ \hline
       7       &    9         & 1 & 2 & $b \bar b$ & $c \bar c$ \\ \hline
       8       &    10        & 1 & 2 & $b \bar b$ & $c \bar c$ \\ \hline
       9       &    9         & 1 & 2 & $c \bar c$ & $b \bar b$ \\ \hline
       10      &    10        & 1 & 2 & $c \bar c$ & $b \bar b$ \\ \hline
       11      &    10        & 2 & 1 & $c \bar c$ & $b \bar b$ \\ \hline
       12      &    9         & 2 & 1 & $c \bar c$ & $b \bar b$ \\ \hline
       13      &    10        & 2 & 1 & $b \bar b$ & $c \bar c$ \\ \hline
       14      &    9         & 2 & 1 & $b \bar b$ & $c \bar c$ \\ \hline
       15      &    6         & 1 & 2 & $b \bar b$ & $c \bar c$ \\ \hline
       16      &    7         & 1 & 2 & $b \bar b$ & $c \bar c$ \\ \hline
       17      &    8         & 1 & 2 & $b \bar b$ & $c \bar c$ \\ \hline
       18      &    6         & 2 & 1 & $c \bar c$ & $b \bar b$ \\ \hline
       19      &    6         & 1 & 2 & $c \bar c$ & $b \bar b$ \\ \hline
\end{tabular}
\begin{tabular}{|c|c|c|c|c|c|} \hline
 $ \alpha  $   & the diagram  & a & b &     1      &     2      \\ \hline
       20      &    7         & 2 & 1 & $b \bar b$ & $c \bar c$ \\ \hline
       21      &    8         & 2 & 1 & $c \bar c$ & $b \bar b$ \\ \hline
       22      &    6         & 2 & 1 & $b \bar b$ & $c \bar c$ \\ \hline
       23      &    12        & 1 & 2 & $b \bar b$ & $c \bar c$ \\ \hline
       24      &    5         & 1 & 2 & $b \bar b$ & $c \bar c$ \\ \hline
       25      &    13        & 2 & 1 & $b \bar b$ & $c \bar c$ \\ \hline
       26      &    12        & 2 & 1 & $b \bar b$ & $c \bar c$ \\ \hline
       27      &    5         & 2 & 1 & $b \bar b$ & $c \bar c$ \\ \hline
       28      &    13        & 1 & 2 & $b \bar b$ & $c \bar c$ \\ \hline
       29      &    12        & 1 & 2 & $c \bar c$ & $b \bar b$ \\ \hline
       30      &    5         & 1 & 2 & $c \bar c$ & $b \bar b$ \\ \hline
       31      &    13        & 2 & 1 & $c \bar c$ & $b \bar b$ \\ \hline
       32      &    12        & 2 & 1 & $c \bar c$ & $b \bar b$ \\ \hline
       33      &    5         & 2 & 1 & $c \bar c$ & $b \bar b$ \\ \hline
       34      &    13        & 1 & 2 & $c \bar c$ & $b \bar b$ \\ \hline
       35      &    4         & 1 & 2 & $b \bar b$ & $c \bar c$ \\ \hline
       36      &    3         & 1 & 2 & $b \bar b$ & $c \bar c$ \\ \hline
       37      &    4         & 1 & 2 & $c \bar c$ & $b \bar b$ \\ \hline
       38      &    3         & 1 & 2 & $c \bar c$ & $b \bar b$ \\ \hline
\end{tabular}
\normalsize
\end{center}
\end{table}

 The matrix element squared for the subprocess
$ gg \rightarrow b \bar b  c \bar c $
is obtained from the formula:
$$ {|M_{gg}|}^2 = (4\pi \alpha_s)^4 \cdot
\sum_{\mu_i ,\lambda_j}^{} \Bigl( \sum_{\alpha ,\beta}^{}
\tilde M_{\alpha}^{*} (\mu_i ,\lambda_j) C_{\alpha \beta}
\tilde M_{\beta} (\mu_i ,\lambda_j) \Bigr),$$
where $\mu_i$--polarization states of initial gluons, $\lambda_j$-- a set
of helicity states of final fermions, $\tilde M_{\gamma}$-- helicity part
of the contribution from the appropriate Feynmann diagram,  $C_{\alpha
\beta}$--
colour matrix calculated by a separate program by using the  formula:
$$ C_{\alpha \beta}=\sum_{i_1 , i_2 , f_1 , f_2 , f_3 , f_4}^{}
C_{\alpha}^{*}( i_1 , i_2 , f_1 , f_2 , f_3 , f_4 ) \cdot
C_{\beta}( i_1 , i_2 , f_1 , f_2 , f_3 , f_4 ),$$
where $i_1 , i_2 $--colour states of initial partons,
$f_1 , f_2 , f_3 , f_4$--colour states of final quarks and
$C_{\gamma}$-- colour part of the appropriate Feynmann diagram.
The cross-section of the subprocess under consideration is obtained
through  Monte-Carlo integration over the phase space
of $b \bar b c \bar c $ particles and averaging over the colour and helicity
states of initial gluons.

An analogous calculation method can also be used for the subprocess
 $q{\bar q}\to b(q_1)+{\bar b}(q_2)+c(q_3)+{\bar c}(q_4)$.

\section{$B_c$-meson production cross-section.}

We confine ourselves to the consideration of the  $S$-wave
 states of $B_c$-mesons.
The underlying assumption of our calculations is that the binding energy
of two quarks, $b$, $\bar c$ is much less than their masses and, hence, heavy
quarks in the bound state $B_c$ are actually on the mass shell. In this case
the 4-momenta $p_b$ and $p_c$ of the quark constituents of $B_c$-meson are
related with the 4-momentum $P$ of $B_c$-meson as follows
\begin{center}
$$  p_b=\frac{m_b}{M}P  \qquad \qquad   p_c=\frac{m_c}{M}P  $$
\end{center}
where $M=m_b+m_c$ is the $B_c$-meson mass.
Hence, the $B_c$-meson production is described by the diagrams of Fig. 1
 by combining two quark lines to meson line and
singling out the states with definite quantum numbers.
The amplitude of $B_c$-meson production can be expressed through
appropriate projection operators and helicity amplitudes
$M_{h{\bar h}}(\lambda_i)$ as follows:
\begin{displaymath}
M(\lambda_i)=\frac{\sqrt {2M}}{\sqrt {2m_b} \sqrt {2m_c}} \Psi (0)
\sum_{h,\bar h,q, \bar q }^{}P_{h,\bar h}M_{h,\bar h,q, \bar q}(\lambda_i)
\delta_{q, \bar q}
\end{displaymath}
where summation is made over the helicity
$h,\bar h$ and colour $q,\bar q$ indices of the quark and antiquark producing
$B_c$-meson. The helicities of remaining fermions are symbolically
denoted through $\lambda_i$. Projection operators $P_{h,\bar h}$
have the following explicit form ($H=h-{\bar h}$)[1,3,12]:

$$ P_{h,\bar h}=\frac{1}{\sqrt 2}{(-1)}^{ {\bar h}-1/2 }\delta_{H,0} $$
for the $S_0$-state and

$$ P_{h,\bar h}=|H|+\frac{1}{\sqrt 2}\delta_{H,0} $$
for the $S_1$-state.

The value of the wave function at the origin $\Psi (0)$ is calculated
in the nonrelativistic potential model and also in the QCD sum rules [13]
and is related with the decay constant $f_{B_c}$ of the pseudoscalar
$B_c(0^{-})$-meson and constant $f_{B^*_c}$ of the vector
$B^*_c(1^{-})$-meson as follows:

$$ \Psi (0)=\sqrt{\frac{M}{12}}f_{B_c},$$ where
$$f_{B_c}=f_{B^*_c}=570{\ \rm MeV}$$

Potentials of various types yield actually the same values of the masses
of low lying states of $B_c(B^*_c)$-mesons. For example, the mass of
pseudoscalar $0^-$-meson (1$S$-state) is $M=6.3$ GeV. In this connection,
when calculating the bound states of $B_c$-mesons of $1S$ levels we take the
 masses of $b$-and $c$-quarks somewhat larger, than those for the production
of free $b{\bar b}c{\bar c}$-quarks and equal to $m_b=4.8$ GeV and $m_c=1.5$
 GeV.
The mass of $2S$ levels is predicted to be $M=6.9$ GeV. In this case the
masses of $b$- and $c$-quarks are taken to be still larger:
$m_b=5.1$ GeV and $m_c=1.8$ GeV.
The value of the wave function at the origin $ \Psi (0)$ for $2S$ states is
$ \Psi (0)= 0.275{\ \rm GeV}^{3/2}$~[14].

The calculation of the colour matrix and integration over the phase space
of final particles is analogous to the case of four free quark production.

The FORTRAN program for the calculation of the matrix element was obtained
by means of slight changes of the program calculating the matrix element
of four free quark production. This program was checked on the Lorentz
invariance with respect to boost along the beam axis and azimuthal invariance
($p_x\to p_y$ and $p_y \to -p_x$).

Table 2 and Fig. 2 show the values of
$\hat \sigma_{gg\rightarrow B_c(B_c^*) \bar b c}$ for several energies
of interacting gluons assuming that $m_b=5.1$ GeV, $m_c=1.5$ GeV
and $\alpha_s=0.2$.

\begin{table}[t]
\caption{Gluon cross-section of $B_c(B_c^*)$-meson production.
(Bracketed is the Monte-Carlo error in the last digit.) }
\begin{center}
\begin{tabular}{|c|c|c|c|c|}    \hline
 $                  $ & 20 GeV & 40 GeV & 100 GeV & 1 TeV \\ \hline
 $\hat\sigma_{B_c}, nb$&$2.93(1)\cdot 10^{-2}$&  $4.0(1)\cdot 10^{-2}$&
 $2.1(1)\cdot 10^{-2}$ &  $2.7(4)\cdot10^{-4} $   \\   \hline
 $ \hat\sigma_{B_c^*}, nb$&$7.16(3)\cdot 10^{-2}$
&$0.105(2)$& $5.6(2) \cdot 10^{-2}$& $6(1)\cdot 10^{-4}$\\   \hline
\end{tabular}
\end{center}
\end{table}

We see that the ratio $\sigma_{B_c^*} /\sigma_{B_c}$ is $\sim 2.5$ for energies
20, 40, 100 GeV and it $\simeq 2$ for  1 TeV, while in
$e^+ e^-$ annihilation
[1-3] the ratio of $\sigma_{B_c^*} /\sigma_{B_c}$ is predicted to be
$\simeq 1.3$. This suggests an idea that the mechanism of the
$B_c(B_c^*)$-meson
production on gluons is in principle different from that in the process
$e^+ e^- \rightarrow B_c(B_c^*) \bar b c$. In order to understand this
difference let us consider the subprocess
$gg \rightarrow B_c(B_c^*)\bar b c$. From Fig. 1 one can see that there are
two types of diagrams. In diagrams
(3), (4), (12), (13) of Fig. 1 the quark-antiquark pair is created
on the leg of quark of a different flavour. We call such diagrams
as  fragmentation ones. As for diagrams
(1), (2), (5)-(11), called recombination
type diagrams, quark-antiquark pairs are created independently.

In [1,2] it was shown, that for the process
$e^+ e^- \to B_c {\bar b}c$ described only by fragmentation
diagrams, for $\frac{M_{B_c}^2}{s} \ll 1$, the mechanism of fragmentation is
realized, and  for the cross-section  equation (1) is right.
Using the same method as in [2], one can easily show, that for the subprocess
$gg\to B_c{\bar b}c$ cross-section, the contribution of fragmentation diagrams
in the regime $\frac{M_{B_c}^2}{s} \ll 1$ can also, as in $e^+e^-$
annihilation, be represented in the factorized form:
\begin{equation}
\frac{d\sigma_{gg \rightarrow B_c(B_c^*) \bar b c}}{dz}=
\sigma_{gg  \rightarrow b \bar b}\cdot D^{(*)}(z),
\end{equation}

where functions $D(z)$ and $D^*(z)$ are the same as in the process
$ e^+ e^- \rightarrow B_c(B_c^*) \bar b c $ and
$\sigma_{gg  \rightarrow b \bar b}$ is the cross-section of the process
$ gg \rightarrow b \bar b $  for the same energy.

For $D(z)$ we will use the following formulas [2,3]:
\begin{eqnarray}
\label{29}
D(z)&=&\frac{8\alpha^2_s|\Psi (0)|^2}{27m^3_c}\frac{(1-r)z(1-z)^2}{(1-rz)^6}
\biggl(
2+(6-12r)z+(5-\frac{62}{3}r+\frac{68}{3}r^2)z^2  \nonumber \\
&+&(-\frac{10}{3}r+\frac{34}{3}r^2-12r^3)z^3+(r^2-2r^3+2r^4)z^4
\biggr)
\end{eqnarray}
for the pseudoscalar $B_c$-meson and
\begin{eqnarray}
\label{30}
D^*(z)&=&\frac{8\alpha^2_s|\Psi (0)|^2}{27m^3_c}\frac{(1-r)z(1-z)^2}{(1-rz)^6}
\biggl(
2+(-2-4r)z+(15-18r+12r^2)z^2 \nonumber \\
&+&(-10r+6r^2-4r^3)z^3+(3r^2-2r^3+2r^4)z^4
\biggr)
\end{eqnarray}
for the vector $B^*_c$-meson.

Unfortunately, in the energy range under consideration the picture of
$B_c(B_c^*)$-meson production in the process
$gg \rightarrow B_c(B_c^*) \bar b c$
turns out to be more complex and
can not be described by simple fragmentation. This can be seen from
Figs. 3a, 4a, 5a, where we show the distribution of
$\frac{d\hat \sigma_{gg\rightarrow B_c \bar b c}}{dz}$ (solid line),
calculated by Monte-Carlo integration of the matrix element
for energies 20, 40, 100 GeV, respectively, in comparison with
the distribution given by (2) (dotted line).

One can see that the cross-sections are much larger than those obtained
from formula (2) for energies 20, 40, 100 GeV.
By dashed lines on these Figures we show the distribution
$\frac{d\hat \sigma^{frag}_{gg\rightarrow B_c \bar b c}}{dz}$,
calculated by means of Monte-Carlo integration of fragmentation contribution.
One can see that the contribution of fragmentation diagrams already
at the energy 40 GeV is described by formula (2) quite well.
Fig. 4d  shows both distributions of
$\frac{d\hat \sigma^{frag}_{gg\rightarrow B_c \bar b c}}{dz}$
(dashed line) and
$\sigma_{gg  \rightarrow b \bar b}\cdot D(z)$ (solid line).
The difference between the last two distributions
is the evidence that nonscaled terms of the order of
$\frac{M^2_{B_c}}{s}$ are still  substantial.
 As is evident from  Fig. 5d nonscaled terms in the fragmentation region
remains also at 100 GeV.

Figs. 6a, 7a,d, 8a,d show
the same distributions but for the case of vector meson production
(at the energies of 20, 40, 100 GeV respectively). From them one can conclude
that the contribution of recombination diagrams to the cross-section
is still larger than in the case of $B_c$-meson production. This explains
the fact that the ratio $\sigma_{B_c^*} /\sigma_{B_c} $ in
$gg$ collisions is different from that in $e^+e^-$ collisions,
where the fragmentation mechanism is dominant.

When comparing distributions over $z$ at different energies one can see
that the form of distributions considerably changes both in the case of
$B_c$- and of $B^*_c$-meson production. The observed dependence of
$\sigma_{B_c^*} /\sigma_{B_c} $ shows that the contribution
of the recombination diagrams decreases  with the energy growth
faster than the contribution of fragmentation diagrams, although  at all the
reasonable energies it remains significantly large.

The fragmentation and recombination contributions have different behavior
over the transverse momentum (see Figs. 4c, 5c). These cross-section
distributions of the gluonic $B_c$-meson production are shown for
energies 40, 100 GeV (solid and dashed lines correspond to  the total
cross-section and its fragmentation part respectively). In the case of
$B^*_c$-meson production analogous distributions are presented in
Figs. 7c, 8c. From them one can conclude that the contribution of
recombination diagrams remains substantial over all the kinematical region
(see also Figs. 4b, 5b for $B_c$ and Figs. 7b, 8b for $B^*_c$, which show
the distributions over $\cos \Theta$,
where $\Theta$ is the angle of $B_c(B^*_c)$
with respect to the beam axis; dashed line is the fragmentation contribution).

Comparing the distributions over $\cos \Theta$ for $B_c$-meson at different
energies one should note that while the $B_c$-meson production not far
from threshold is actually uniform, at higher energies its production
occurs in the region of small angles with respect
to the direction of initial gluons. Analogous note is also valid for
the vector meson production.

The comparison of the  distributions
$\frac{d\hat \sigma_{gg\rightarrow B_c \bar b c}}{d\cos \Theta}$ and
$\frac{d\hat \sigma_{gg\rightarrow B_c^* \bar b c}}{d\cos \Theta}$
at low energy shows that for the same energy the distribution over
$\cos \Theta$ for $B_c$ is more uniform than that for the $B_c^*$.

\section{$B_c(B_c^*)$-meson hadronic production cross-section.}

In the parton model the total cross-section for the
$B_c(B_c^*)$-meson production is expressed through the subprocess cross-section
in the following way:
\begin{equation}
 \sigma_{tot} (s) =
\int \limits_{ (2m_b +2m_c)^2 }^{s} \frac{d \hat s}{s}
\int \limits_{-1+\frac{\hat s}{s}}^{ 1-\frac{\hat s}{s}} \frac{dx}{x^*}
\Bigl(\sum_{i,j} f^i_a (x_1) f^j_b (x_2)\cdot \sigma_{ij}(\hat s) \Bigr),
\end{equation}

where $\hat s$---invariant mass squared of interacting partons;

$x_1,x_2$---portions of parton momenta in the appropriate protons;

   $x=x_1 - x_2 $;
\ $x^*=\sqrt{x^2+4\hat s/s}$;

$f^i_a (x_1)$,$ f^j_b (x_2)$---parton structure functions in interacting
hadrons;
$\sigma_{ij}(\hat s)$---the subprocess
$ij\rightarrow B_c \bar b c$ cross-section.

The gluon structure functions $G(x,Q^2)$ are as follows~[15]:

\begin{displaymath}
x\cdot G(x,Q^2)=K(S)\cdot \exp{\Bigl(12 \sqrt{S \cdot ln(1/x)/B}
\ \Bigr)}\cdot {(1-x)}^6,
\end{displaymath}

where

\begin{displaymath}
\begin{array}{l}
t_0=ln(Q_0^2/\Lambda ^2),\\
t=ln(Q^2/\Lambda ^2),\\
S=ln(t/t_0),\\
B=33-2N_f,\\
\end{array}
\end{displaymath}
\begin{displaymath}
\begin{array}{c}
K(S)=50.36(\exp(S)-0.957)\exp(-7.597\sqrt{S}),\\
 Q_0^2=5{\ \rm GeV}^2 ,\  \Lambda=200{\ \rm MeV},\  N_f=5.\\
\end{array}
\end{displaymath}

The structure functions of valent quarks $u$, $d$ in the proton
 ($\bar u$, $\bar d$ in the antiproton) have the following form:
\begin{displaymath}
\begin{array}{c}
x\cdot U(x,Q^2)=K'(S)\cdot \exp{\Bigl(4 \sqrt{2} \sqrt{S \cdot ln(1/x)/B}
\ \Bigr)}\cdot x^{0.65} \cdot {(1-x)}^3,
\\
x\cdot D(x,Q^2)=0.5\cdot (1-x)\cdot x\cdot U(x,Q^2),

\end{array}
\end{displaymath}
where $K'(S)=2\sqrt{S}\exp{(-1.5S)}$.

The structure functions of sea quarks $\bar u$, $\bar d$ in the proton
 ($u$, $d$ in the antiproton) are equal to:
\begin{displaymath}
x\cdot \bar U(x,Q^2)=x\cdot \bar D(x,Q^2)
=0.5 \cdot \sqrt{S \cdot ln(1/x)/B}\cdot x \cdot G(x,Q^2).
\end{displaymath}

In Tab. 3  we present the values of total production cross-sections
of  $B_c$-meson $S$-levels  for different energies.
\begin{table}[t]
\caption{Hadronic production cross-section of $B_c(B_c^*)$.(Bracketed is the
Monte-Carlo error in the last digit.)  }
\begin{center}
\begin{tabular}{|c|c|c|c|c|}    \hline
 $n^{2S+1}L_J$ & $1{}^1S_0$ & $1{}^3S_1$ & $2{}^1S_0$ & $2{}^3S_1$ \\ \hline
 $\sigma(40\ \rm GeV)$, $nb\cdot 10^{-5}$ &  $1.07(1)$ &  $ 4.44(6)$
& $0.0846(5)$ &  $0.387(5) $   \\   \hline
 $\sigma(100\ \rm GeV)$, $nb\cdot 10^{-3}$ &  $6.18(6)$ &
$ 17.7(2)$   & $0.873(8)$
 &  $2.45(3) $   \\   \hline
 $\sigma(1.8\ \rm TeV)$, $nb$ &  $12.2(3)$ &  $ 32.2(2)$   & $2.7(1)$
 &  $ 6.8(4)  $  \\ \hline
$\sigma(16\ \rm TeV)$, $nb\cdot 10^{2}$ &  $1.96(8)$ &  $ 4.99(6)$   &
$0.42(3)$
 &  $ 1.1(1) $\\  \hline
\end{tabular}
\end{center}
\end{table}

The energy of 40 GeV is close to the c.m.s. energy for the fixed target
experiments
at HERA, calculations at $\sqrt{s}=1.8$ TeV are made bearing in mind the
experiments at Tevatron and, at last, the energy $\sqrt{s}=16$ TeV
corresponds to the conditions of $pp$-experiment at LHC.
The energy dependence of the cross-section summed also over $\bar B_c$
is shown in Fig. 9.

{}From Tab. 3 it follows that for $\sqrt{s}=40$ GeV the summed meson production
cross-section $\sigma_{sum}$ is $\sim 10^{-4}$ of the total $b \bar b$
production cross-section which makes the study a very complicated task in
 these experiments. One should note that in this case
one cannot use only the $gg\to B_c{\bar b}c$ contributions and we
take into account also the $q{\bar q}\to B_c{\bar b}c$ one.

There are experiments at Tevatron and LHC that will have a real possibility
to discover the hadronic $B_c$ production, because in the latter case
$\frac{\sigma_{sum}}{\sigma_{b \bar b}}$ is $\sim 10^{-2}$.
For this reason we give for the last two colliders the most interesting
distributions of $1{}^1S_0$ and $1{}^3S_1$ production cross-sections.
Note, that as our calculations have show, the cross-section
at the considered energies is completely determined by gluon interaction
(the suppression of quark-antiquark contribution is $\sim 10^{-2}$).

Figs.10 and 11 present the distributions for $1{}^1S_0$ (dashed line) and
$1{}^3S_1$ (solid line) at the energy of 1.8 TeV of colliding hadrons.
Distributions $\frac{d \sigma_{B_c(B_c^*)}}{dx}$ (see Fig. 11b)
show  that the particles are created in the central region,
 the cross-section dominates in the interval from -0.3 to 0.3.

{}From Fig. 10b it follows that the mean $B_c$-meson transverse momentum
is not large and has the value of the order of 6 GeV. Figs. 10c, 11d, where
the distributions over  $b$- and $c$-quark transverse momenta are presented,
show that mean transverse momenta of remaining free quarks is actually
the same as that of $B_c$-meson.

{}From the distributions of Fig. 11a  over the angle between  the directions
of motions of $B_c$-meson and $c$-quark one can see that in most cases
the $c$-quark moves in the same direction as $B_c$-meson.

It is interesting to note that the distribution over
the invariant mass of three
final particles $M_{{B_c}\bar b c}$ both for vector and pseudoscalar
mesons (Fig. 10a) shows that the cross-section is saturated in the
region of $M_{{B_c}\bar b c}$ from the threshold up to the 100 GeV and that
the mean value of $M_{{B_c}\bar b c}$ is of the order of 30 GeV.

The behavior of the distributions at the energy of colliding hadrons of 16 TeV
is not  much different from those considered above. This can be seen
from Figs. 12 and 13 where the same distributions as in Figs. 10 and 11
are presented but for the energy of 16 TeV.

\section*{Conclusion}

In this work we presented the calculation results  for $B_c(B_c^*)$-meson
hadronic production  and its radial excitations production.
The calculations are made
in the  QCD perturbation theory in the forth order over the
strong coupling constant $\alpha_s^4$.

The successive analysis of various diagram contributions shows that they
may be divided into two types: fragmentation and recombination ones.
The diagrams of the first class describe the process of $b$-quark
 fragmentation
 into  $B_c(B_c^*)$-meson. Such  process is well described
on the basis of $B_c(B_c^*)$ production in $e^+e^-$ annihilation.
The differential cross-section of fragmentation contribution has simple
factorized form (2). Repeating the arguments of the works, devoted to the case
of $e^+e^-$ annihilation, we have confirmed
the existence of such factorization
in two gluon $B_c(B_c^*)$-meson production.
The results of our computer calculations for the fragmentation contribution
coincide with factorized result (2).
Then we showed that the fragmentation contribution in contrast to the existing
point of view [16] is not dominant even in the region of large transverse
momenta.

The basic contribution to the $B_c(B_c^*)$ production is related with
diagrams of recombination type. The contribution of some of them falls
with the energy increase by the  factor of $\frac{1}{\hat s}$
faster than others explaining thus the strong dependence of $B_c(B_c^*)$
spectra on $\hat s$ (in the considered interval of $\hat s$).
One should note that in contrast to the fragmentation contribution, where the
ratio $\frac{\sigma_{B_c*}}{\sigma_{B_c}}$ is 1.3, recombination diagrams give
the ratio $\frac{\sigma_{B_c*}}{\sigma_{B_c}}\sim 2.5$. Such a ratio is
consistent with naive counting of spin degrees of freedom.

The considered diagrams of QCD perturbation theory are of the forth order
with respect to the strong coupling constant $\alpha_s$. This causes
strong dependence on the definite choice of $Q^2$ in the argument of
$\alpha_s(Q^2)$. The choice of $Q^2$ has to be  defined by a typical
virtuality in the production process. The analysis shows that this virtuality
is large only in contributions, which fall faster than
$\frac{1}{\hat s}$. In other cases, including fragmentation contribution,
this virtuality is not large and is of the order of $4m_bm_c$.
For this reason in cross-section estimation we used the value $\alpha_s=0.2$.
Using, for example, the quantity $\alpha_s(\hat s)$ leads to a 7 fold
decrease of the $B_c(B_c^*)$-meson cross-section. However, in this case
it does not explain the disagreement between our predictions and those
from ref. [16] for  $B_c(B_c^*)$ total production cross-section.
In [16] the same diagrams as in the present work are calculated.
Both we and [16] predict the contribution $q{\bar q}\to B_c{\bar b}c$
to be small. However, in [16] the definition of $Q^2$ in the strong
coupling constant $\alpha_s(Q^2)$ is absent.
Probably, the authors of [16] have used  the definition $\alpha_s({\hat s})$.
But even in this case, according to our opinion,
the $B_c$ and $B^*_c$ production cross-section
obtained in ref. [16] is too small: it is even smaller than the
contribution of the fragmentation component alone [17].

\vspace*{0.5cm}
In conclusion authors express their gratitude to S.S.Gerstein,
V.V.Kiselev and A.V.Tkabladze for useful discussions. The work of
M.V.Shevlyagin was supported, in part, by Russian Fund of Fundamental
Investigations (Number 93-02-14456).

\newpage
\section*{Figure captions}
\begin{itemize}
\item[Fig. 1] Types of diagrams of the four free quarks gluonic production.
\item[Fig. 2] The cross-section of the subprocesses
 $gg \rightarrow B_c \bar b c$ (blank marker) and
$gg \rightarrow B_c^* \bar b c$ (black marker ) as function of the gluonic
energy. The dependence
$2\cdot 10^{-3} \cdot \sigma_{gg \rightarrow b \bar b}$  is presented
 by solid line.
\item[Fig. 3] The cross-section distributions of the gluonic $B_c$-meson
 production for energy of 20 GeV (see Section 2) over: \\
a)  $z$ (dotted line --- fragmentation mechanism
prediction ); \\
b)  $\cos \Theta$; \\
c)  the transverse momentum.
\item[Fig. 4]{The cross-section distributions of the gluonic $B_c$-meson
 production (solid lines) and the  distributions of fragmentation part of this
production (dashed lines) for energy of 40 GeV (see Section 2) over: \\
a)  $z$ (dotted line --- fragmentation mechanism
prediction ); \\
b)  $\cos \Theta$; \\
c)  the transverse momentum; \\
d)  $z$ for the fragmentation contribution (dashed line) and for
the fragmentation mechanism prediction (solid line).}
\item[Fig. 5]{The cross-section distributions of the gluonic $B_c$-meson
 production (solid lines) and the  distributions of fragmentation part of this
production (dashed lines) for energy of 100 GeV (see Section 2) over: \\
a) $z$ (dotted line --- fragmentation mechanism
prediction ); \\
b)  $\cos \Theta$; \\
c)  the transverse momentum; \\
d)  $z$ for the fragmentation contribution (dashed line) and for
the fragmentation mechanism prediction (solid line).}
\item[Fig. 6]{The cross-section distributions of the gluonic $B_c^*$-meson
 production for energy of 20 GeV (see Section 2) over: \\
a)  $z$ (dotted line --- fragmentation mechanism
prediction ); \\
b) $\cos \Theta$; \\
c) the transverse momentum. }
\item[Fig. 7]{The cross-section distributions of the gluonic $B_c^*$-meson
 production (solid lines) and the  distributions of fragmentation part of this
production (dashed lines) for energy of 40 GeV (see Section 2) over: \\
a)  $z$ (dotted line --- fragmentation mechanism
prediction ); \\
b)  $\cos \Theta$; \\
c) the transverse momentum; \\
d)  $z$ for the fragmentation contribution (dashed line) and for
the fragmentation mechanism prediction (solid line).}
\item[Fig. 8]{The cross-section distributions of the gluonic $B_c^*$-meson
 production (solid lines) and the  distributions of fragmentation part of this
production (dashed lines) for energy of 100 GeV (see Section 2) over: \\
a) $z$ (dotted line --- fragmentation mechanism
prediction ); \\
b) $\cos \Theta$; \\
c)  the transverse momentum; \\
d)  $z$ for the fragmentation contribution (dashed line) and for
the fragmentation mechanism prediction (solid line).}
\item[Fig. 9]{The total  $B_c$-meson hadronic production (blank marker) in $nb$
 and the  $b \bar b$-pair hadronic production in $\mu b$
(black marker).  }
\item[Fig. 10]{The cross-section distributions of the  $B_c$-meson
 production (solid lines) and the $B_c^*$-meson production (dashed lines)
in the process $ p \bar p \rightarrow B_c \bar b c+ X$
 for energy of 1.8 TeV (see Section 3) over: \\
a)  the energy of the interacting partons; \\
b)  the transverse momentum of $B_c(B_c^*)$; \\
c)  the transverse momentum of $\bar b$-quark; \\
d)  the rapidity. }
\item[Fig. 11]{The cross-section distributions of the  $B_c$-meson
 production (solid lines) and the $B_c^*$-meson production (dashed lines)
in the process $ p \bar p \rightarrow B_c \bar b c+ X$
 for energy of  1.8 TeV (see Section 3) over: \\
a) the angle cosine between the directions of
motions of $B_c(B_c^*)$-meson and $c$-quark; \\
b) $x$; \\
c)  the modulus of the $B_c(B_c^*)$-meson momentum; \\
d)  the transverse momentum of $ c$-quark.}
\item[Fig. 12]{The cross-section distributions of the  $B_c$-meson
 production (solid lines) and the $B_c^*$-meson production (dashed lines)
in the process $ p p \rightarrow B_c \bar b c+ X$
 for energy of 16 TeV (see Section 3) over: \\
a)  the energy of the interacting partons; \\
b)  the transverse momentum of $B_c(B_c^*)$; \\
c)  the transverse momentum of $\bar b$-quark; \\
d)  the rapidity. }
\item[Fig. 13]{The cross-section distributions of the  $B_c$-meson
 production (solid lines) and the $B_c^*$-meson production (dashed lines)
in the process $ p p \rightarrow B_c \bar b c+ X$
 for energy of 16 TeV (see Section 3) over: \\
a) the angle cosine between the directions of
motions of $B_c(B_c^*)$-meson and $c$-quark; \\
b)  $x$; \\
c)  the module of the $B_c(B_c^*)$-meson momentum; \\
d)  the transverse momentum of $ c$-quark.}
\end{itemize}

\end{document}